\documentclass[11pt]{article}
\usepackage{amssymb}
\usepackage[numbers,sort&compress]{natbib}
\usepackage{amsmath,amsfonts,graphicx,epsfig}
\usepackage{bm}
\usepackage{caption}
\usepackage{hyperref}
\usepackage{cleveref}
\usepackage{xcolor}

\def\xv{\vec{x}}
\def\bea{\begin{eqnarray}}
\def\eea{\end{eqnarray}}
\def\be{\begin{equation}}
\def\ee{\end{equation}}
\def\ba{\begin{array}}
\def\ea{\end{array}}

\def\kp{{k_{\rm peak}}}

\def\R{{\mathcal R}}

\def\F{{\mathcal F}}

\def\P{{\mathcal P}}

\def\C{{\mathcal C}}

\begin{document}

\setlength\arraycolsep{2pt}

\renewcommand{\theequation}{\arabic{section}.\arabic{equation}}
\setcounter{page}{1}


\begin{center}

\vskip 1.0 cm

{\LARGE  \bf The role of non-gaussianities in Primordial Black Hole formation}

\vskip 1.0cm

{\Large
Vicente Atal \& Cristiano Germani
}

\vskip 0.5cm

{\it 
Departament de F\'isica Fonamental i Institut de Ci\'encies del Cosmos, Universitat de Barcelona, Mart\'i i Franqu\'es 1, 08028 Barcelona, Spain.
}

\vskip 1.5cm

\end{center}

\begin{abstract} 
We re-analyse current single-field inflationary models related to primordial black holes formation. We do so by taking into account recent developments on the estimations of their abundances and the influence of non-gaussianities. We show that, for all of them, the gaussian approximation, which is typically used to estimate the primordial black holes abundances, fails. However, in the case in which the inflaton potential has an inflection point, the contribution of non-gaussianities is only perturbative.
Finally, we infer that only models featuring an inflection point in the inflationary potential, might predict, with a very good approximation, the desired abundances by the sole use of the gaussian statistics.
\end{abstract}

\setcounter{equation}{0}
\section{Introduction}

LIGO detection of black holes mergers \cite{Abbott:2016blz} have renewed the interest in the hypothesis that the dark matter, or some substantial fraction of it, might be composed by primordial black holes (PBHs) \cite{Ivanov:1994pa,Bird:2016dcv, Sasaki:2016jop, Carr:2016drx,sibi}. Within this hypothesis, PBHs can be generated, among other mechanisms, as a consequence of high {\it non-linear} peaks in the primordial distribution of density perturbations \cite{Carr:1974nx}. While at the Cosmic Microwave Background Radiation (CMB) scales the amplitudes of the curvature perturbations are too small to generate a significant amount of PBHs, there is currently no hard bound on their amplitudes at small scales, leaving open the possibility of having a large fraction of the Dark Matter (DM) in the form of PBHs. 

The method to determine PBH abundances has been recently extensively revisited \cite{ilia,ilia2} (see also, \cite{Yoo:2018kvb}). In \cite{ilia,ilia2} it has been shown that the abundances are sensitive to the shape of the power spectrum, and that a proper account of this might drastically change earlier predictions. A caveat might be there though. In the work of \cite{ilia}, and in most of the earlier works, it was assumed an exact gaussian distribution for the amplitudes of density perturbations. This assumption might seem too strong as rare non-linear events (associated to the PBH production) sit in the tail of the probability distribution of the density perturbations.

The question of whether the gaussian estimation of abundances is correct has been raised in \cite{Bullock:1996at, Saito:2008em,Bugaev:2013vba,Young:2013oia,Young:2014oea, Young:2015cyn, Pattison:2017mbe, Franciolini:2018vbk} and, in \cite{Franciolini:2018vbk}, a method to check whether or not non-gaussianities are relevant has been provided. Applying this method to the correct statistical variable describing the PBHs abundances, we will show that the gaussian estimation receives large corrections from the inflationary bispectrum in all current single-field models of PBH production. The models considered in this work are all the single-field models constructed so far that are able to generate a large peak in the power spectrum.

The earliest attempt to connect single-field models to PBHs formation appeared in \cite{Yokoyama:1998pt}. However, there, the predicted CMB spectrum was ruled out just after the measurements of the Planck satellite \cite{Planck:2013jfk}.  The first models consistent with current CMB data were designed to have an inflection point in the inflationary potential. From this class of models we will consider the one in Germani et al. \cite{Germani:2017bcs}, which optimises the model of Garcia-Bellido et al. \cite{GB}. These models generate a relatively small peak in the power spectrum. A more efficient way to generate a large peak is to consider a potential with a local maximum. There, the peak amplitude is at least an order of magnitude larger than the one related to an inflection point. Inspired by Higgs inflation \cite{bond}, Ezquiaga et al. \cite{Ezquiaga:2017fvi}\footnote{Here, the slow-roll approximation was used in the regime in which the ultra-slow-roll one should be instead used. For a similar analysis with realistic runnings of the Higgs parameters and within the slow-roll regime, the authors of \cite{Bezrukov:2017dyv} found a negligible abundance of PBHs.}, Kannike et al. \cite{Kannike:2017bxn}, Ballesteros et al. \cite{Ballesteros:2017fsr} and R\"as\"anen et al. \cite{Rasanen:2018fom}\footnote{Here, the authors assumed Planck mass remnants after the evaporation of very small PBHs. In this paper we will not consider this case.} considered a non-minimally coupled scalar with logarithmically running coupling constants. It was also realised that potentials with a local maximum can be also found in UV inspired settings such as stringy axions (\"Ozsoy et al. \cite{Ozsoy:2018flq}), alpha-attractors in supergravity (Dalianis et al. \cite{Dalianis:2018frf}) and specific settings in string theory (Cicoli et al. \cite{Cicoli:2018asa})\footnote{On the listed models, we will only consider the parameters which predicts the smallest non-gaussian signal.}. In this work we will consider for each model the parameters which predicts the smallest non-gaussian signal. We also assume that the process of PBH formation happens during radiation era, which is the case in a `standard' reheating history.

\section{Abundances and the impact of non-gaussianities}

\subsection{The over-density and distribution}

Inflation generates, at super-horizon scales, the following metric perturbations on a Friedmann-Robertson-Walker (FRW) geometry
\be\label{R}
ds^2\simeq-dt^2+a(t)^2 e^{2\R(\xv)}d\vec{x}\cdot d\vec{x}\ ,
\ee
where $\R$ is the so-called curvature perturbation, $a(t)$ the scale factor and the horizon size is defined as $R_H\equiv1/\dot a$. Initial conditions for numerical simulations of primordial black hole formation are imposed at super-horizon scales and in real space \cite{ilia2}. Thus, the minimal amplitude (threshold or critical value) that a perturbation must have to trigger gravitational collapse into a black hole, is typically given at those scales.

It is clear from \eqref{R} that the curvature perturbation is defined up to an arbitrary constant, so it is meaningless to define an absolute threshold value for $\R$. The physical quantity is the density contrast in real space \cite{Young:2014ana,ilia,Yoo:2018kvb}, $\Delta(\xv,t)\equiv\delta\rho(\xv,t) / \rho(t)$, where $\rho(t)$ is the background energy density and $\delta\rho(\xv,t)$ its perturbation. At linear order and at super-horizon scales, $\Delta(\xv,t)$ is related to $\R(\xv)$ as\footnote{The linear approximation is good enough to make our point in this paper. In the full non-linear treatment, the critical value for collapse changes by a factor ranging from $1$ to $2$ with respect to the linear case \cite{Yoo:2018kvb}.}
\be\label{relation}
\Delta(\xv,t) \simeq \frac{4}{9}\frac{1}{a^2H^2} \nabla^2 \R(\xv)\ ,
\ee 
or, in Fourier space
\be
\Delta_k\simeq  \frac{4}{9}\frac{k^2}{a^2H^2} \R_k\ .
\ee
The amplitudes of the curvature perturbations $\R_k$ are statistically distributed accordingly to the specific inflationary model considered. Assuming $\R_k$ to be gaussian, since PBHs are rarely formed, the over-densities related to PBHs formations are, with a very good approximation, spherically symmetric \cite{Bardeen:1985tr}. Under this assumption, one can define the averaged energy density $\bar\Delta$ as \cite{ilia}
\be\label{eq:two_pf_delta}
\bar\Delta(r,t)=\frac{{\cal F}_0}{a^2 H^2}\psi(r)\ ,
\ee
which is parameterised by the statistically distributed central value amplitude ${\cal F}_0$. In \eqref{eq:two_pf_delta}, the shape of the density perturbation is described by the function
\be
\psi(r)\equiv\frac{\int dk k^2\frac{\sin\left(kr\right)}{kr}P_\Delta(k,t)}{\int dk k^2 P_\Delta(k,t)} \ ,
\ee
where $P_\Delta(k,t)$ is given by
\be
\left(2\pi\right)^3P_\Delta(k,t)\,\delta(k,k')\equiv \langle\Delta(k,t)\Delta(k',t)\rangle \ .
\ee
It has been shown in \cite{Bardeen:1985tr} that the number density of peaks, $n$, is a function of the variable $\nu\equiv \frac{{\cal F}_0}{\tilde\sigma_{\Delta}}$, where, at super-horizon scales,
\be
\tilde\sigma_{\Delta}^2(r)\equiv (a H)^4 \sigma_\Delta^2(r,t)\ ,
\ee
with
\be
\sigma_\Delta^2(r,t)\equiv\int \frac{k^2}{2\pi^2}dk P_{\Delta}(k,t)\ .
\ee
Finally, given a threshold ${\cal F}_0^c$ determining whether an initial perturbation eventually collapses into a black hole, the abundances of PBHs are proportional to the integral $\int_{\nu_c}^ \infty n(\nu)d\nu$, where $\nu_c\equiv \frac{{\cal F}_0^c}{\tilde\sigma_{\Delta}}$.  

\subsection{Threshold}

In order to find the threshold ${\cal F}_0^c$ we need to define few other quantities \cite{ilia2}: The over-mass generated by the averaged curvature perturbations is, at super-horizon scales
\be
\frac{\delta M}{M}\equiv \delta(r,t)\simeq \frac{3}{r^3}\int_0^r \bar\Delta(r',t)r'^2 dr' \ .
\ee
From this, we can define the compaction function $\C\simeq 2 \delta M/\left(a(t)r\right)$ and the point $r_m$ in which $\C(r)$ is maximal. The critical value $\delta_{c}\equiv\delta (r_m,t_m)$ associated to ${\cal F}_0^c$, where $a(t_m)H(t_m)r_m=1$, has been found to range from $0.41$ to $0.67$ \cite{Harada:2013epa,Harada:2015yda}. It is easy to show \cite{ilia2} that $\delta_c=3\bar\Delta(r_m,t_m)$ and therefore 
\be
{\cal F}_0^c=\frac{\delta_c}{3\psi(r_m)r_m^2}\ .
\ee
In other words, apart from the mild dependence on the exact numerical evolution of the initial density fluctuation, the critical value ${\cal F}_0^c$ can always be semi-analytically calculated given a primordial power spectrum.

\subsection{Window functions}

As discussed in the previous section, the number density of PBHs depends on the point in which the compaction function has a local maximum. Therefore, different local maxima of the compaction function are related to different values for the threshold. Moreover, at the threshold, only the portion of the over-density within a radius $r_{\rm max}\lesssim 2 r_m$ from the centre of the over-density contributes to the PBH formation \cite{ilia2}. Suppose $r_m^{\rm min}$ is the smallest radius where the compaction function has a maximum. One may ask what is the threshold such that a portion of the over-density with radius {\it larger} than $2 r_m^{\rm min}$ participates to the PBH formation. In order to answer this question, we need to smooth-out scales smaller than $2 r_m^{\rm min}$ and search for the subsequent local maximum of $\cal C$. This filtering process would also change the statistics of the peaks.

As explained for example in \cite{Bardeen:1985tr}, the smoothing-out of small scales can be done by applying a window function. In particular the most convenient one is a Gaussian since it does not have spurious oscillations neither in physical nor momentum space (see e.g. \cite{Kolb:1990vq}). Small scales are  then smoothed-out at a scale $k_{\rm cut}$ by the replacement
\be
\Delta_k(t)\rightarrow e^{-\frac{k^2}{2 k_{\rm cut}^2}}\Delta_k(t)\ ,
\ee
and $r_m=r_m(k_{\rm cut})$.

Given a specific inflationary model, the end of inflation is the moment ($t_{\rm end}$) in which the whole power spectrum is at super-horizon scales. Therefore, it is natural to fix the initial conditions for the evolution of the primordial density fluctuations at $t_{\rm end}$. The power spectrum has therefore an inherited cut-off at the scale $k_{\rm end}=a(t_{\rm end})H(t_{\rm end})$. What we will show later on is that the abundances of PBHs are larger when the smoothing-out momentum is smaller than $k_{\rm end}$.
 
\subsection{Non-Gaussian contribution}

Our discussion so far requires that the deviations from gaussianity of the statistical variable $\Delta_k$ are small enough to be neglected.

In \cite{Franciolini:2018vbk}, a criterion was given to estimate the contribution of the skewness of $\Delta_k$ in the calculation of the PBHs abundances. There, it was shown that in order to trust the gaussian estimate, the combination
\be\label{eq:E3}
E_3\equiv 0.19\ {\cal F}_0^3 \frac{\tilde S_3}{\tilde\sigma_{\Delta}^2}\ ,
\ee
where
\be
S_3\equiv a^6 H^6\frac{\langle\Delta(0)^3\rangle}{\tilde\sigma_\Delta^4}=\frac{a^6 H^6}{\tilde\sigma_{\Delta}^4}\int \frac{d^3 k_1}{2\pi^3} \int \frac{d^3 k_2}{2\pi^3} \int \frac{d^3 k_3}{2\pi^3}\, \langle \Delta_{k_1}\Delta_{k_2}\Delta_{k_3}  \rangle  \ ,
\ee
must be, in absolute value, smaller than $1$.\footnote{In principle the full non-linear relation between $\Delta_k$ and $\R_k$ should be used. However, assuming that the higher order correlations of $\R_k$ are small, one can safely test the non-gaussian contributions just by using the linear relation.} The reason is that the ratio of the non-gaussian corrected PBHs density at formation over the total density ($\beta^{\rm NG}$), is related to the abundances calculated by the sole use of gaussian statistics ($\beta^{\rm G}$) by
\be\label{e3}
\beta^{\rm NG}=e^{E_3}\beta^{\rm G}\ .
\ee

Note however that, in the case in which $E_3$ is significant, the right-hand side of equation \eqref{e3} cannot be calculated by the gaussian statistics. The reason is that the non-gaussianities will also generically modify the profile of the over-density.

\section{Application to known single-field models related to PBHs formation}\label{sec:NG} 

For all the single-field models introduced earlier, we are going to show that the power spectrum has a peak generated when the inflaton is very close, in field space, to the local maxima (or inflection point) of its own potential. This will allows us to make some generic predictions for the amplitude and shape of the non-gaussianities.

Models of inflation\footnote{Here and for the rest of the paper we mean models of single-field inflation.} are mainly characterised by two (geometrical) slow-roll parameters, $\epsilon=-\frac{\dot H}{H^2}$ and $\epsilon_2=\frac{\dot \epsilon}{H\epsilon}$, where $H$
 is the Hubble constant. So far, inflationary scenarios related to the production of PBHs are typically either a cascade of quasi-slow-roll phases (QSR), see for example \cite{GB}, where $\epsilon\ll 1$ and $-3<\epsilon_2<0$, or interpolate between the following  three phases of roughly constant $\epsilon_2$:  
 \begin{itemize}
 \item 1) {\it slow-roll phase} (SR): $\epsilon , \epsilon_2 \ll 1$ and the perturbations related to the cosmic microwave background (CMB) are generated. 
 \item 2) {\it constant-roll phase} (CR) or {\it ultra-slow-roll} (USR): $\epsilon\ll 1$ and $\epsilon_2 \leq -6 $ (where the equality is for USR). This is the phase in which a large peak in the power spectrum is produced. The reason is that here curvature perturbations grow exponentially at horizon crossing and at super-horizon scales.
 \item 3) {\it graceful exit} (GE): $\epsilon\ll 1$ and $\epsilon_2\gg \epsilon$. In this phase super-horizon curvature perturbations are frozen and inflation evolves into its end ($\epsilon\sim 1$). 
 \end{itemize}
 The inflaton potential related to a cascade of QSR trajectory does not have any extrema. On the contrary, in the case of  trajectories interpolating from phase 1 to 3  the inflaton potential has an inflection point if phase 2 is USR, or a false minimum if of CR type.

The scenarios of a cascade of QSR phases can only generate a small fraction of DM in terms of PBHs\footnote{Unless accretion mechanisms are invoked \cite{GB}.}, and non-gaussianities are bound to be small \cite{Maldacena:2002vr}.
On the other hand, in the interpolating trajectory, a larger fraction can be achieved as there, curvature perturbations pass trough a phase of exponential growth. Indeed, the main difference between the QSR and CR/USR cases is that in the latter the curvature perturbations are not constant at super-horizon scales. Thus, here we focus on the classes of models with an interpolating trajectory, since they optimally generate a large peak in the power spectrum. 

Generalising \cite{Cai:2017bxr} to the constant-roll case\footnote{In \cite{Cai:2017bxr} only a transition from USR to GE was studied.}, we will show that there is a specific combination of the slow-roll parameters that is kept approximately fixed during the transition from USR/CR to GE, provided that the transition from 2) to 3) happens in a sufficiently short field range. This combination is the mass parameter $\alpha$ appearing in the Mukhanov-Sasaki equation, which describes the evolution of the variable $v_k\equiv z \R$
\be\label{eq:MS}
v_k''+\left(k^2-\left(\alpha^2-\frac{1}{4}\right)\tau^{-2}\right)v_k=0\ .
\ee
In \eqref{eq:MS} $z=a\sqrt{2\epsilon}$, $\tau$ the conformal time and prime denotes derivative with respect to it. 

Phase 2) is characterised by a constant value of $|\epsilon_2|$ (and a negligible value of $\epsilon$). Then one finds that $\alpha^2\simeq \frac{9}{4}+\frac{3}{2}\epsilon_2+\frac{1}{4}\epsilon_2^2$. 

It is easy to see that, in the case in which $\epsilon\ll\epsilon_2$ and $\epsilon_2$ constant, the transformation $\epsilon_2 \rightarrow -6-\epsilon_2$ keeps $\alpha^2$ invariant \cite{kinney,Tzirakis:2007bf,Morse:2018kda}. Then, because $\alpha^2$ is constant in the transition between phase 2 and 3, this duality necessarily implies that the values of $\epsilon_2$ at the constant roll ($\epsilon_2^{cr}$) and at the {\it beginning} of the graceful exit phase when curvatures perturbations get frozen ($\epsilon_2^{ge}$), are related trough the relation
\be
\epsilon_2^{cr} = -6-\epsilon^{ge}_2 \ .
\ee
The implications of this duality for the background and linear perturbations have been already extensively studied in \cite{kinney,Tzirakis:2007bf,Morse:2018kda}. Extending these results to non-gaussianities, we will show that the duality between the USR/CR and GE phases persist at the level of the (third order) non-linear perturbations: the dominant contribution to $f_{\rm NL}$ stays constant from phase 2 to phase 3 and it is of local shape. In particular $f_{\rm NL}$ can be of $\mathcal{O}(1)$ whenever the local maxima of the potential is steep enough.

In a FRW universe with metric $ds^2=-dt^2+a^2d\vec{x}^2$, the dynamics of a canonical scalar field $\phi(t)$ is governed by the following equations
\begin{align}\label{eq:back1}
&\ddot{\phi}+3H\dot{\phi}+\frac{dV}{d\phi} = 0  \ , \\
&3 H^2 = \frac{1}{2}\dot{\phi}^2+V \ , \label{eq:back2}
\end{align} 
where $H=\dot{a}/a$, dot denotes cosmic time, and the reduced Planck mass $M_{pl}=1$. At zeroth order approximation, the power spectrum of curvature perturbations is inversely proportional to the inflaton velocity $\dot\phi$. Therefore, in a quasi-DeSitter space (inflation), one needs to go from a ``high'' velocity (the SR phase related to CMB) to a lower one in a very short time. In other words, one needs a large negative acceleration to form a peak in the power spectrum. 

From the scalar equation we have
\be
\Big|\frac{\ddot\phi}{3 H\dot\phi}\Big|=\Big|1+\frac{dV/d\phi}{3 H\dot\phi}\Big|\equiv \Big|1-\Delta_V\Big|\ .
\ee     
A deceleration is given for $\Delta_V<1$. During SR, $\Delta_V\sim 1$ and so there is approximately no deceleration. A QSR phase would be for $0<\Delta_V<1$, a USR for $\Delta_V=0$ and CR for $\Delta_V<0$, all of them will make $|\dot\phi|$ to decrease. In USR the inflaton decelerates at the inflection point thanks to the Hubble friction and then directly evolves into the graceful exit. The case in which phase 2 is instead a CR corresponds to the case in which the sign of the  potential gradient changes. Here, the inflaton climbs up a potential barrier before reaching a maximum and then entering into a graceful exit.

During the climb up, curvature perturbations grow exponentially due to the exponential decrease of the field velocity. One might then naively say that the peak of the power spectrum is generated when the field velocity reaches its minimum (in modulus).
However the dynamics are slightly more complicated: At the crossing horizon time ($t_*$), perturbations get a power spectrum inversely proportional to $\epsilon(t_*)$ and so exponentially grow with $t_*$. In addition, after horizon crossing, the perturbations instead of being frozen as in the SR phase, grow exponentially. In USR the combination of the two effects generates a scale invariant spectrum \cite{Kinney:2005vj}. In CR however the spectrum is red \cite{Martin:2012pe}. In other words, the prolongated secular growth that larger scales experiences,  `wins' over the increasing amplitude at horizon crossing of the smaller scales. 

Since the velocity of the inflaton is exponentially suppressed during CR, the peak of the power spectrum must nevertheless lie very close (in field space) to the top of the potential. This is necessary in order to overcome the barrier and evolve into a graceful exit phase. In the following we will see that this feature is shared by all existing examples of single-field inflation related to PBHs formation.

\subsection{Duality in the transition between USR/CR and GE}\label{sec:review}

\begin{figure}[h!]
\begin{center}
\includegraphics[scale=0.5]{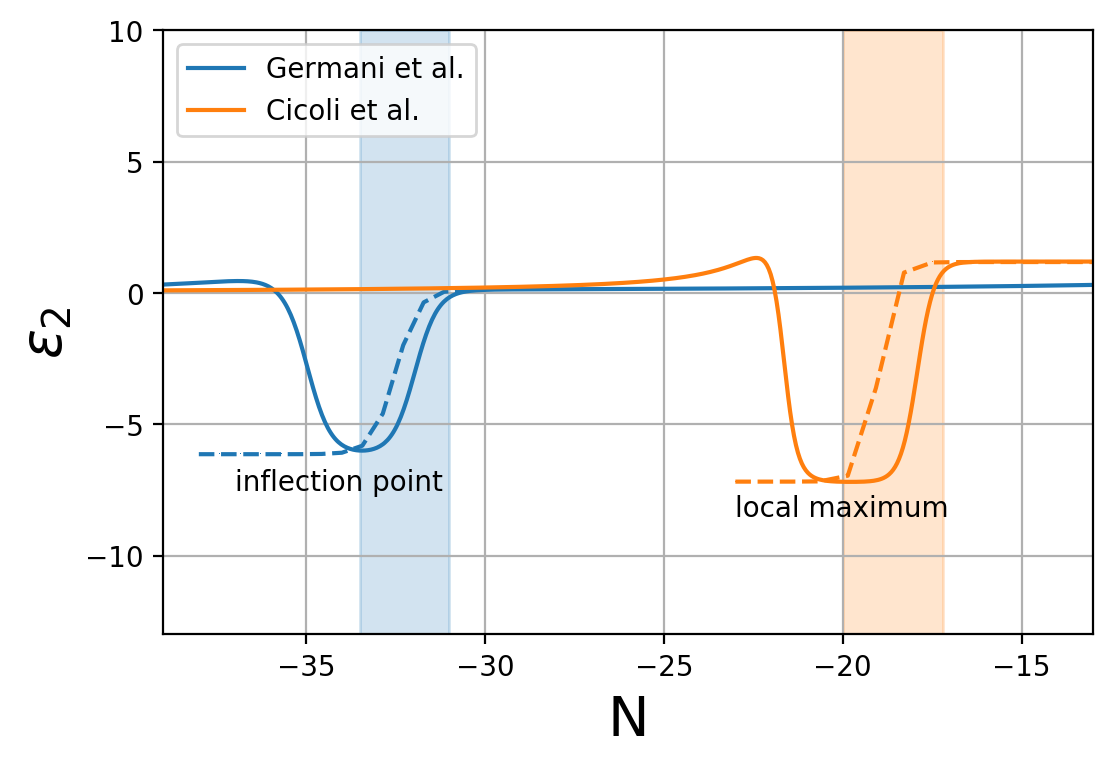}
\hspace{5cm}
\includegraphics[scale=0.5]{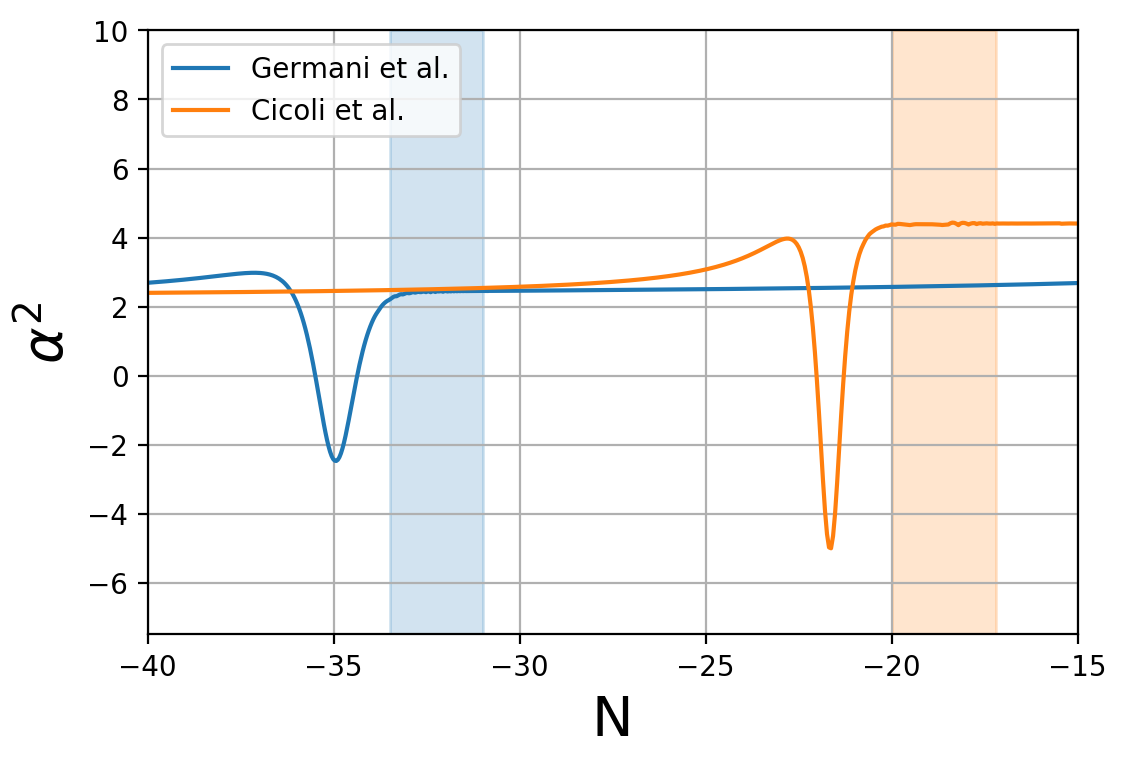}
\includegraphics[scale=0.5]{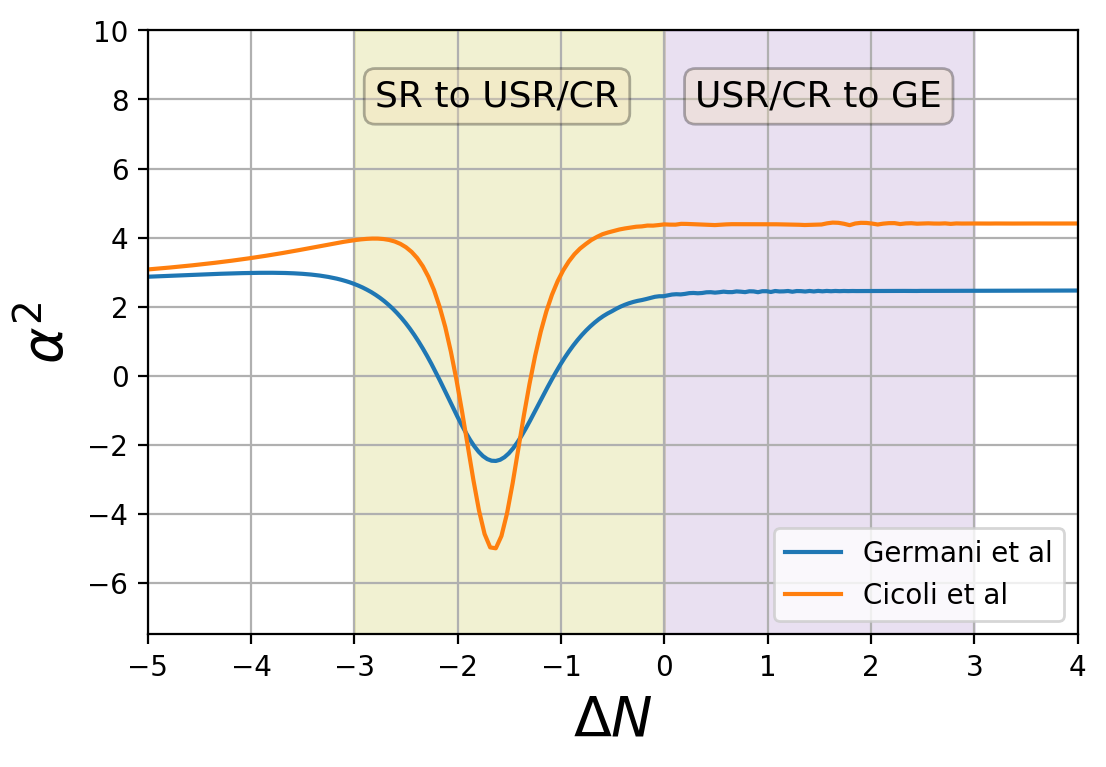}
\caption{\textit{Top)} Evolution of $\epsilon_2$ for the two representative models with an inflection point \cite{Germani:2017bcs}, and with a local maximum \cite{Cicoli:2018asa}. The dashed lines are the predictions given by eq. (\ref{eq:eps2_dual}), where we have slightly shifted the time so that the correspondence can be better appreciated (otherwise the two curves superimpose).\textit{Bottom) left)} Evolution of the parameter $\alpha^2$. The shaded regions corresponds to the time when USR/CR transitions to GE. During the transition, $\alpha^2$ is constant (for all different possible values of $\alpha^2$). \textit{Bottom) right)} detail of the left plot.}
\label{fig:eps2_vs_N}
\end{center}
\end{figure}

In the upper panel of Fig. \ref{fig:eps2_vs_N} we show the evolution of $\epsilon_2$ for two representative models featuring an inflection point \cite{Germani:2017bcs} and a local maximum \cite{Cicoli:2018asa} in the inflationary potential. We have numerically checked that all the models share the same qualitative features for $\epsilon_2$ and $\alpha^2$. In the bottom panel we show the evolution of $\alpha^2$ for the same models. We see that $\alpha^2$ is indeed invariant in the USR/CR to GE transition (and thus for any value of $\epsilon_2$ in the USR/CR phase). 

In all models that we have studied, the transition from CR to GE happens around and very close to the top of the potential, and the peak of the power spectrum is related to scales that exit the horizon during the transition (more precisely, at the beginning, when the value of $|\epsilon_2|$ is maximised). Thus, all the interesting dynamics are captured by expanding the potential around the maximum or the inflection point, depending upon the nature of the potential. 

If we want to describe a potential having a maximum (or an inflection point), we need at least to expand it up to second order 
\begin{equation}
V=V_0 + \sqrt{2\epsilon_V}V_0(\phi-\phi_{0})+\frac{\eta_V}{2}V_0(\phi-\phi_{0})^2 +\ldots\label{exp}
\end{equation}
where we define
\begin{equation}
 \epsilon_V = \frac{1}{2}\left(\frac{V'}{V}\right)^2\Big|_{\phi=\phi_0}  \quad,\quad  \eta_V= \frac{V^{''}}{V}\Big|_{\phi=\phi_0} \ ,
\end{equation}
and $\phi_0$ is either the position of the local maximum or the inflection point. For models having a local maximum $\sqrt{\epsilon_V}=0$ and $\eta_V \neq 0$. As shown in \cite{Cai:2017bxr}, the second derivative of the potential is related to the mass 
parameter of the Mukhanov-Sasaki equation $\alpha^2=9/4+\frac{3}{2}\epsilon_2+\frac{1}{4}\epsilon_2^2+\frac{\dot{\epsilon_2}}{2H}+\epsilon$. Irrespective on whether the inflaton is in slow-roll or not, it holds that
\be\label{eq:nu_cont}
\alpha^2-9/4=-\frac{V^{''}}{H^2}+\mathcal{O}(\epsilon) \ ,
\ee 
\noindent meaning that the parameter $\alpha^2$ is constant provided the expansion (\ref{exp}) holds. For models passing trough a local maximum one can have $\frac{V^{''}}{H^2} \simeq 3\eta_V \sim \mathcal{O}(1)$, such that $\alpha^2$ is a constant different from $9/4$. For models passing through an inflection point $\alpha^2\simeq 9/4$. There, a small departure from $9/4$ is given by the third order derivative of the potential. Interestingly, Eq. (\ref{eq:nu_cont}) allows us to solve the dynamics of the background without having to solve explicitly the equations of motion. Solving directly (\ref{eq:nu_cont}) for $\epsilon_2$ we find that (whenever $\epsilon_2$ is larger than $\epsilon$), it evolves according to the simple relation (at leading order in $\epsilon$)
\be\label{eq:eps2_dual}
\epsilon_2=-3+\sqrt{9 - 12 \eta_V} \tanh\left[\frac{\sqrt{9 - 12 \eta_V} \left(N-N_0\right)}{2}\right] \ ,
\ee
where $N_0$ is an integration constant that can be fixed by the inflationary initial conditions.

If the third derivative of the potential, in the expansion \eqref{exp}, is never important in the trajectory from the CR to the beginning of the GE phase, the solution \eqref{eq:eps2_dual} interpolates between $\epsilon^{cr}_2=-3-\sqrt{9-12\eta_V}$, at the CR phase, and $\epsilon_2^{ge}=-6-\epsilon_2^{cr}=-3+\sqrt{9-12\eta_V}$, at the beginning of the GE phase\footnote{Afterwards $\epsilon_2$ might no longer follow \eqref{eq:eps2_dual}.}.  Note that because we are near a local maximum, $\eta_V$ is negative and needs not to be small so that $\epsilon_2^{ge}$ might be $\mathcal{O}(1)$ (and for non-negligible amount of efoldings).

As already mentioned, the evolution of $\epsilon_2$ in \eqref{eq:eps2_dual}, from phase 2 to phase 3, is in very good agreement with what we have numerically found in all the models that we have studied:
For the potentials related to a CR phase, (\ref{eq:eps2_dual}) differs by less than 2 percent compared to the values of $\epsilon_2^{cr}$ and $\epsilon_{2}^{ge}$ that we have numerically found. In the USR case, $\epsilon_2^{usr}$ is very accurate also, but the predicted value for $\epsilon_2^{ge}$ is less. In fact, for potentials having $\eta_V=0$, the analytical estimation for $\epsilon_2^{ge}$ in (\ref{eq:eps2_dual}) would give exactly $0$. However, in this case there would not be a graceful exit. Although the expansion in \eqref{exp} is still valid, since the second derivative of the potential is exactly zero, the third order term of the expansion will determine the graceful exit phase. This term contributes to the value of $\epsilon_2^{ge}$ with a non vanishing (but still small) contribution. For example, in the model with an inflection point that we have considered, we find $\epsilon_2^{ge}\sim 0.1$ which allows us to still approximately use (\ref{eq:eps2_dual}), as done in \cite{Cai:2017bxr}. 

Finally, we would like to point out that \eqref{eq:eps2_dual} is the time dependent version of the already mentioned duality existing between CR/USR and SR models \cite{Tzirakis:2007bf,Morse:2018kda}.  This duality transforms $\alpha$ into (-$\alpha$), keeping $\alpha^2$ invariant. Under this duality, the power spectrum of the variable $v_k$ is the same in the phase 2 and 3 (the shift $\alpha$ to ($-\alpha$) brings an irrelevant phase factor into the mode equations \cite{Morse:2018kda}). Note however that the power spectrum of $\R$ is not the same in the phases $2$ and $3$. The reason is that the latter is defined by an additional suppression factor of $\epsilon$, which is obviously not constant between the phases 2 and 3.

\subsection{Bispectrum} \label{ng}

In this section we show that the duality discussed previously also holds at the level of the bispectrum, provided the background evolution is given by Eq. (\ref{eq:eps2_dual}). This generalises the results found for the background and linear perturbations \cite{kinney,Tzirakis:2007bf,Morse:2018kda}\footnote{The discussion on non-gaussianities can be already found in the context of USR in \cite{Cai:2017bxr}.}.

At third order, the action for $\R$ is the one given by \cite{Maldacena:2002vr}
\begin{multline}\label{eq:s3}
S_3 = \int d^4x\, \left(a^3 \epsilon^2 \R \dot{\R}^2+ a \epsilon^2 (\partial \R)^2-2 a \epsilon \dot{\R}(\partial \R)(\partial \chi)+ \frac{a^3 \epsilon}{2} \dot{\epsilon_2} \R^2 \dot{\R} + \frac{\epsilon}{2a} (\partial \R)(\partial \chi)\partial^2 \chi \right. \\ \left.+  \frac{\epsilon}{2a} (\partial^2 \R)(\partial \chi)^2 + 2f(\R)\frac{\delta \mathcal{L}}{\delta \R} \right) 
\end{multline}
where
\begin{equation}
 \partial^2 \chi = a^2 \epsilon \dot{\R} \quad,\quad \frac{\delta \mathcal{L}}{\delta \R} = a( \partial^2 \dot{\chi} + H  \partial^2 \chi - \epsilon  \partial^2 \R )
\end{equation}
and 
\begin{equation}\label{eq:fielred}
 f(\R) = \frac{\epsilon_2}{4} \R^2 +2\R\frac{\dot\R}{H}+\ldots\ .
\end{equation}
In \eqref{eq:fielred} ``$\ldots$'' denote spatial gradients that can be neglected at super-horizon scales. The terms proportional to $f(\R)$ can be removed from the third order action by performing a field redefinition $\R \rightarrow \R_n+ f(\R_n)$. 

Whenever $\R$ is frozen at super-horizon scales, as in the GE phase, the relation between the correlation functions of $\R_n$ and $\R$ only depends on the first term of ($\ref{eq:fielred}$): 
\be\label{eq:zeta_zetaN}
\langle \R(k_1) \R(k_2) \R(k_3) \rangle = \langle \R_n(k_1) \R_n(k_2) \R_n(k_3) \rangle +  \frac{\epsilon_2}{2} \left(\langle \R(k_1) \R(k_2) \rangle \langle \R(k_1) \R(k_3) \rangle + \text{cyc. perm.} \right) \ ,
\ee
and do not further evolve.

Having a functional form for $\epsilon_2$, we can easily determine the evolution of $\R$ and the time at which a given mode freezes so to quantitatively define the beginning of the GE phase. 

At super-horizon scales one has
\be\label{eq:zeta_super}
\R_{k\rightarrow 0}=c_1+c_2\int \frac{dN}{a^3\epsilon}\ ,
\ee
\noindent where $c_1$ and $c_2$ are two constant. The functional form for $\epsilon$ can be obtained by integrating Eq. (\ref{eq:eps2_dual}) and it is
\begin{align}
\epsilon&=\epsilon_0 \exp\left[ \int \epsilon_2 dN \right] \nonumber \\
&=\epsilon_0\, e^{-3 (N-N_0)} \cosh\left[\frac{\sqrt{9 - 12 \eta_V} \left(N-N_0\right)}{2} \right]^2  \ ,
\end{align}
where $\epsilon_0$ is the value of $\epsilon$ at $N_0$. The time dependent part of $\R$ is then 
\begin{align}
\int \frac{dN}{a^3\epsilon}\propto& \int \text{sech}\left[\frac{\sqrt{9 - 12 \eta_V} \left(N-N_0\right)}{2}\right]^2 dN \nonumber \\
\propto & \, \tanh\left[\frac{\sqrt{9 - 12 \eta_V} \left(N-N_0\right)}{2}\right]-\tanh\left[\frac{\sqrt{9 - 12 \eta_V} \left(N_{\star}-N_0\right)}{2}\right] \nonumber \\
\propto & \, \epsilon_2(N)-\epsilon_2(N_{\star}) \label{eq:zeta_super_2}\ .
\end{align}
\noindent where $N_{\star}$ is the time when a particular mode exits the horizon. From here, by looking at (\ref{eq:eps2_dual}), we see that the curvature perturbations approach exponentially a constant. Note that $\epsilon_2$ might continue to evolve afterwards, but in the graceful exit phase the integrand in \eqref{eq:zeta_super} will exponentially decay in time so that the curvature perturbations will stabilise to the value reached at the beginning of phase 3.

Let us note that while $\epsilon_2(N)$ in the distant past is also approaching a constant (given by $\epsilon_2^{cr}$ or $\epsilon_2^{usr}$), the integral in (\ref{eq:zeta_super_2}) is actually growing exponentially: in the distant past ($N_{\star}<N\ll N_0$) we can expand (\ref{eq:zeta_super_2}) as
\begin{align}
\int \frac{dN}{a^3\epsilon} \propto\, & \epsilon_2(N)-\epsilon_2(N_{\star})  \nonumber \\
\propto\, &  e^{\sqrt{9-12 \eta_V} \Delta N } + \mathcal{O}(e^{\frac{\sqrt{9-12 \eta_V}}{2} \Delta N })^3 + \mathcal{O}(e^{\sqrt{9-12 \eta_V} \Delta N_{\star} })^2\ ,
\end{align}
where $\Delta N = N-N_0<-1$ and $\Delta N_{\star} = N_\star-N_0$. At the crossing horizon $\Delta N=\Delta N_\star$ therefore, as time evolves, the mode function grows as $\R \propto e^{\sqrt{9-12 \eta_V} N} \propto  e^{(|\epsilon_2|-3) N}$. This growing exactly matches the known growth of a pure USR or CR phase, as it should. 

On the other hand, in the distant future ($\Delta N_{\star} \ll 0$ and $\Delta N \gg 0$) the expansion reads
\begin{align}
\int \frac{dN}{a^3\epsilon} \propto\, & \epsilon_2(N)-\epsilon_2(N_{\star})   \nonumber \\
\propto\, & C + e^{-\sqrt{9-12 \eta_V} \Delta N } + \mathcal{O}(e^{-\frac{\sqrt{9-12 \eta_V}}{2} \Delta N })^3 + \mathcal{O}(e^{\sqrt{9-12 \eta_V} \Delta N_{\star} })^2
\end{align}
\noindent where $C$ is a constant. Thus at late time, the time dependent part of the mode function is, as expected and discussed before, a constant.

Equipped by this knowledge, we can now show that the CR/USR and GE bispectra are intimately related via the dual transformation of $\epsilon_2$ discussed before: during GE, $\R$ is constant at super-horizon scales and, because $\epsilon_2\gg \epsilon$ and the operators proportional to $V'''$ are assumed to be negligible with respect to those proportional to $\epsilon_2$, the dominant contribution to the bispectrum will be given only by the field redefinition term. The reason is that all the other couplings are proportional to either $\epsilon$ or $V'''$ \cite{Cai:2017bxr}\footnote{The only non-explicit suppression is in the term $\dot{\epsilon_2} \R^2 \dot{\R}$ of (\ref{eq:s3}). In \cite{Cai:2017bxr}, it was noted that this operator can be re-written, in the flat gauge, as $V^{'''} \delta\phi^3$.}.

Then, the bispectrum is of the following local form 
\be\label{3pt}
\langle \R(k_1) \R(k_2) \R(k_3) \rangle = \frac{\epsilon_2(N_{fr})}{2} \left( \langle \R(k_1) \R(k_2) \langle \R(k_1) \R(k_3) \rangle + \text{cyc. perm.} \right)+{\cal O}(\epsilon^2) \ ,
\ee
where $N_{fr}$ represents the moment (in e-foldings) in which curvature perturbations are frozen and the ${\cal O}(\epsilon^2)$ come from the correlations of $\R_n$s. 

As we have already shown, in our case, $\epsilon(N_{fr})\simeq\epsilon_2^{ge}$ for \textit{all} modes exiting the horizon in the period in which the background is given by \eqref{eq:eps2_dual}. 
Defining $P(k_1)$ as $\langle \R(k_1) \R(k_2) \rangle \equiv (2\pi)^3\delta^3(k_1+k_2)P(k_1)$, the bispectrum $B(k_1,k_2,k_3)$ as $\langle \R(k_1) \R(k_2) \R(k_3) \rangle \equiv (2\pi)^3\delta^3(k_1+k_2+k_3)B(k_1,k2,k3)$, and $f_{\rm NL}$ as
\begin{equation}
f_{\rm NL}(k_1,k_2,k_3)\equiv\frac{5}{6}\frac{B(k_1,k_2,k_3)}{P(k_1)P(k_2)+P(k_1)P(k_3)+P(k_2)P(k_3)} \ ,
\end{equation}
from \eqref{3pt} we get
\begin{align}\label{eq:fnl_ge}
f_{\text{NL}}&=\frac{5}{12}\epsilon_2^{ge} \\ 
&\simeq\frac{5}{12}\left(-3+\sqrt{9-12\eta_V}\right) \label{eq:fnl_etav}\ .
\end{align}
We would like here to stress that the result \eqref{eq:fnl_ge} is valid for {\it all scales} exiting the horizon from phase 2 to phase 3 (and therefore also valid for the scale where the peak in the power spectrum is). In this sense there is a duality between the (third order) perturbations generated during the USR/CR and GE phases: although the bispectrum varies with scale in the transition, the parameter $f_{\text{NL}}$ is all the way constant. 

In the case of USR however, although \eqref{eq:fnl_ge} is still valid, the dual description of $\epsilon_2^{ge}$ in terms of $\eta_V$, i.e. equation \eqref{eq:fnl_etav}, would imply $f_{\rm NL}$ to vanish. This is due to the fact that $V''=0$. Therefore, $\epsilon_2^{ge}$ is fully determined by the value of $V'''$ at the maximum. 

Note that \ref{eq:fnl_ge} does not follow the consistency relations found in \cite{Maldacena:2002vr,Namjoo:2012aa, Martin:2012pe, Bravo:2017wyw}. The reason is that in our case we need to calculate the non-gaussianities at the exit of a USR/CR phase rather than during it.

\subsection{Numerical results}
\begin{figure}[h!]
\begin{center}
\includegraphics[scale=0.6]{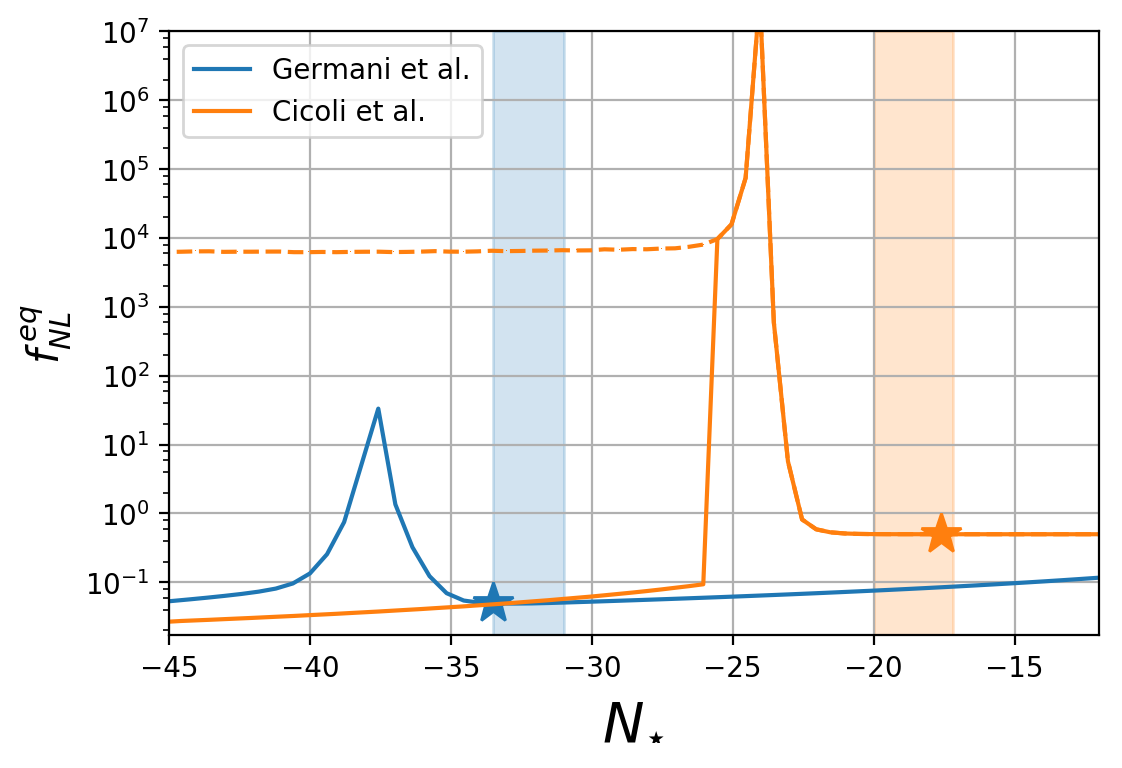} \hspace{5cm}
\includegraphics[scale=0.5]{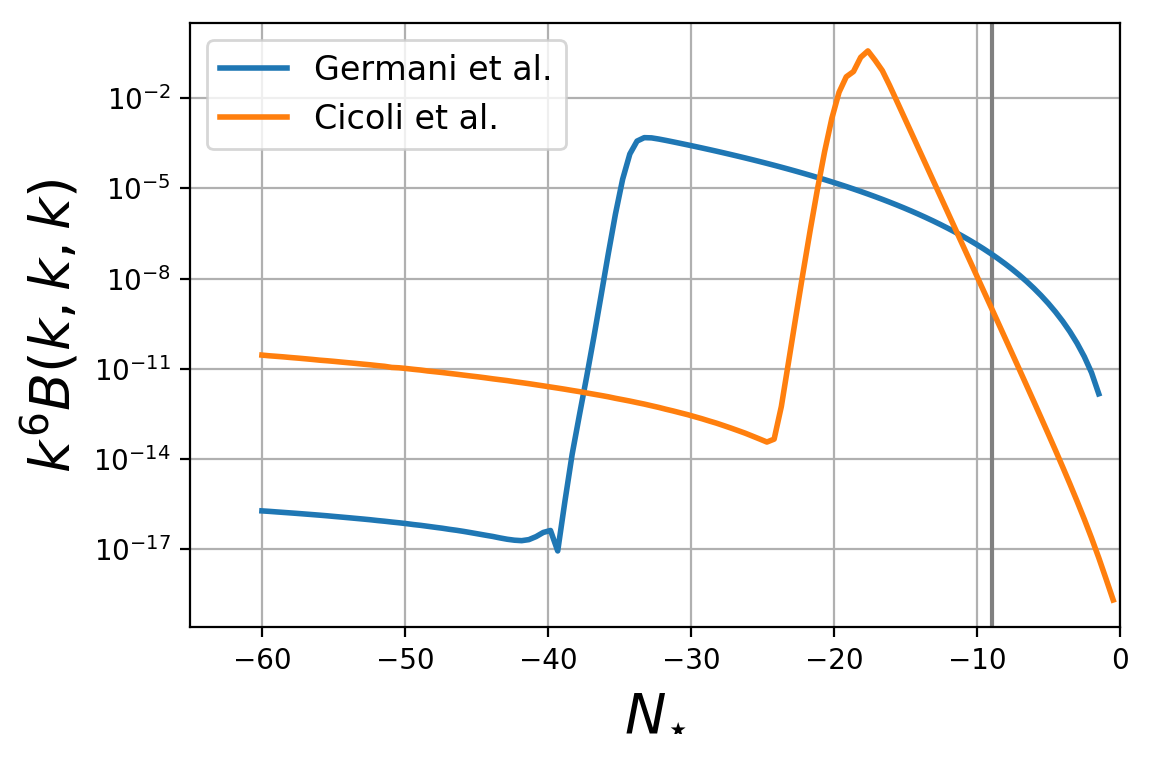}
\includegraphics[scale=0.5]{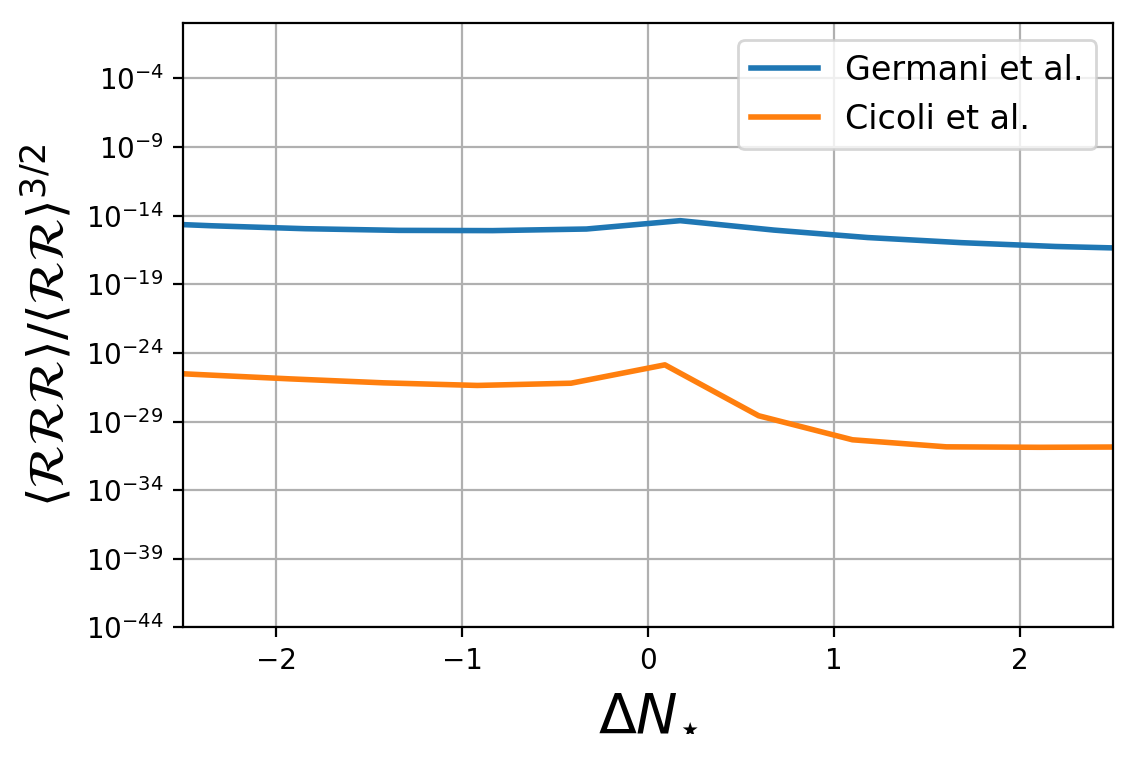}
\caption{\textit{Top)} Parameter $f_{\rm NL}$ as a function of scale. A star represents the scale at which a given model has a peak in the power spectrum. The dashed line represents the actual output of the code if we let the very large scales evolve until the end of inflation. These modes, that exited the horizon much before the USR phase, exhibit a grow in their mode function at that time that we attribute to secular numerical effects. \textit{Bottom) left)} The bispectrum $k^6B(k,k,k)$, in the equilateral configuration. The peak in $f_{\rm NL}$ correspond to scales where the bispectrum is minimized.  \textit{Bottom) right)}  $\langle \R \R \R \rangle / \langle\R \R\rangle^{3/2}$ at the scales of the peak of $f_{\rm NL}$. The ratio is much smaller that one, meaning that the theory is well within the perturbative regime.}
\label{fig:fnls}
\end{center}
\end{figure}

In the following we will corroborate numerically the above estimations by evaluating the three-point function using the public code PyTransport \cite{Dias:2016rjq,Mulryne:2016mzv}\footnote{The code does not handle non-explicit functions for the inflaton potential. Therefore, we are not able to numerically compute $f_{\rm NL}$ in the model of Ballesteros et al.. However the background around the peak of the power spectrum is also well described by eq. (\ref{eq:eps2_dual}), so we expect $f_{\rm NL}$ to follow the same qualitative behaviour of the rest of the models.}. In the top panel of Fig. \ref{fig:fnls} we show the value of $f_{\rm NL}$ in the equilateral configuration as a function of wavelength of the perturbations. For clarity of the figure we only show the USR model of Germani et al. and have selected the CR model of Cicoli et al.. In these examples, we see, as predicted in the previous section, that during the evolution from USR/CR to the GE (in particular during the generation of the peak of the power spectrum) $f_{\rm NL}$ is constant\footnote{Or varies slightly in the model passing through a USR (inflection point potential).} (the shaded region in Fig. \ref{fig:fnls} corresponds to the time when the transition happens). Additionally, at those scales $f_{\rm NL}$ is also independent on the triangle configuration. In other words the three-point function at those scales, and therefore also at the peak of the power spectrum, is of the local shape.

We also note a very large peak in $f_{\rm NL}$ around the minimum of the bispectrum corresponding to the scales leaving the horizon during the transition from SR to USR/CR. This peak  corresponds to scales where the power spectrum and the bispectrum reaches a minimum (as can be seen in the bottom panel for the case of the bispectrum). Although one could be worried that perturbation theory is broken at those scales, that it not actually the case. A test on how good is perturbation theory can be done by checking whether $\langle \R \R \R \rangle / \langle\R \R\rangle^{3/2}\ll 1$ (see e.g. \cite{Senatore:2010jy}). In the bottom right panel of Fig. \ref{fig:fnls} we show that this bound is indeed respected.

\begin{figure}[h!]
\begin{center}
\includegraphics[scale=0.7]{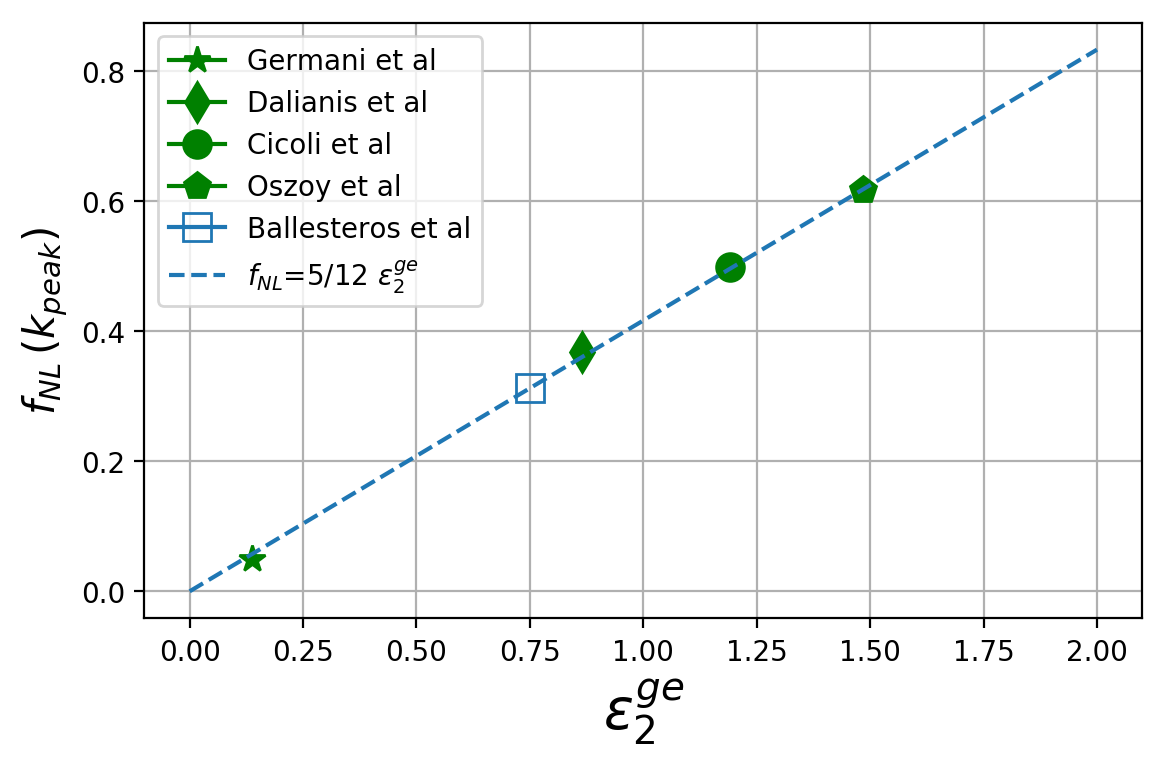}
\caption{$f_{\rm NL}$ at the scale of the peak of the power spectrum, for all models considered in this paper. At this scale, the value of $f_{\rm NL}$ is independent on the triangle configuration, meaning that the shape of the three-point function is of the local shape. The data points are given by the numerical estimates, while the dashed line is the analytical result. For Ballesteros et al., we show where the analytical prediction lies.}
\label{fig:fnl_vs_eps2}
\end{center}
\end{figure}

In Fig. \ref{fig:fnl_vs_eps2}  we show the amplitude of $f_{\rm NL}$ at the scale of the peak of the power spectrum as a function of $\epsilon_2^{ge}$. We see that the estimation given in (\ref{eq:fnl_ge}) works very well. As we already mentioned, the relation between $\epsilon_2$ and $\eta_V$ fails when $\eta_V\sim 0$. For the inflection point model (\ref{eq:fnl_etav}) would predict $f_{\rm NL}= 0$, while we obtain $f_{\rm NL}\simeq 0.05$ for the particular model we have studied. As we have already argued, this difference can be attributed to the presence of a non-negligible third derivative in the inflationary potential at the infection point.

\subsection{Consequences on current models of inflation}

In Table \ref{table:E3_data_nmaj} we show the values for $\F_0r_m^2$, $\nu_c$ and $E_3$ for all the models under consideration and taking into account the indetermination of $\delta_c$. 

We have numerically checked that the larger abundances (minimum value of $\nu_c>1$) of PBHs are obtained by cutting-off the power spectrum at the peak ($k=\kp$). In this case, $\kp r_m \sim 2$.
\begin{center}
  \begin{tabular}{ | c | c | c | c | c |}
    \hline  
    & $\F_0r_m^2$  & $\nu_c$ & $E_3$ & $f_{\rm NL}$     \\ \hline
    Germani et al. \cite{Germani:2017bcs} & $\left[1.2 - 1.9\right]$ & $\left[14 - 22\right]$ &  $\left[4 - 19\right]$ & 0.05 \\ \hline
     Dalianis et al. \cite{Dalianis:2018frf} & $[1.1 - 1.9]$ & $[6 - 10]$ & $\left[7 - 29\right]$ & 0.4 \\ \hline
    Ballesteros et al. \cite{Ballesteros:2017fsr} & $[1.1 - 1.9]$ & $[3 - 5]$ & $\left[2 - 7\right]$ & 0.3 \\ \hline   
    \"{O}szoy et al. \cite{Ozsoy:2018flq} & $[1.1 - 1.8]$ & $[3 - 4]$ & $\left[2 - 9\right]$ & 0.6  \\ \hline
    Cicoli et al. \cite{Cicoli:2018asa} & $[1.2 - 1.9]$ & $[5 - 7]$ & $\left[5 - 21\right]$ & 0.5\\ \hline
  \end{tabular}
\captionof{table}{The values for $\F_0r_m^2$, $\nu_c$, $E_3$ and $f_{\rm NL}$ for the five different models considered.}
  \label{table:E3_data_nmaj}
\end{center}

For all models we find $E_3>1$ and so all the predictions for the abundances are sensitive to the third-order momenta, and possibly higher. However, a special attention should be given to the USR case, where $f_{\rm NL}$ is the smallest. 

The fact that the non-gaussian contribution is ``large" in the USR model of \cite{Germani:2017bcs} should not be a surprise. Indeed, it was already known that the model of \cite{Germani:2017bcs} only generates a relatively small amplitude of the power spectrum. Thus there, chances of producing a PBH are related to the very far tail of the probability distribution of ${\cal F}_0$ which in turn, is extremely sensitive to small non-gaussianities. Nevertheless, by looking at the ratio $\nu_c^2/(2 E_3)$, we see that the non-gaussianities enter here only perturbatively and thus we expect that higher momenta will be subdominant. 

In addition, we also note that a hypothetical model with an inflection point and a smaller $V'''$ with respect to the one of \cite{Germani:2017bcs}, could be well able to predict the right abundances of PBHs by the sole use of the gaussian statistics as, in this case, $f_{\rm NL}$ would be smaller.

 \section{Analytical explanation of the results found}\label{sec:PowerLaw}

\begin{figure}[h!]
\begin{center}
\includegraphics[scale=0.8]{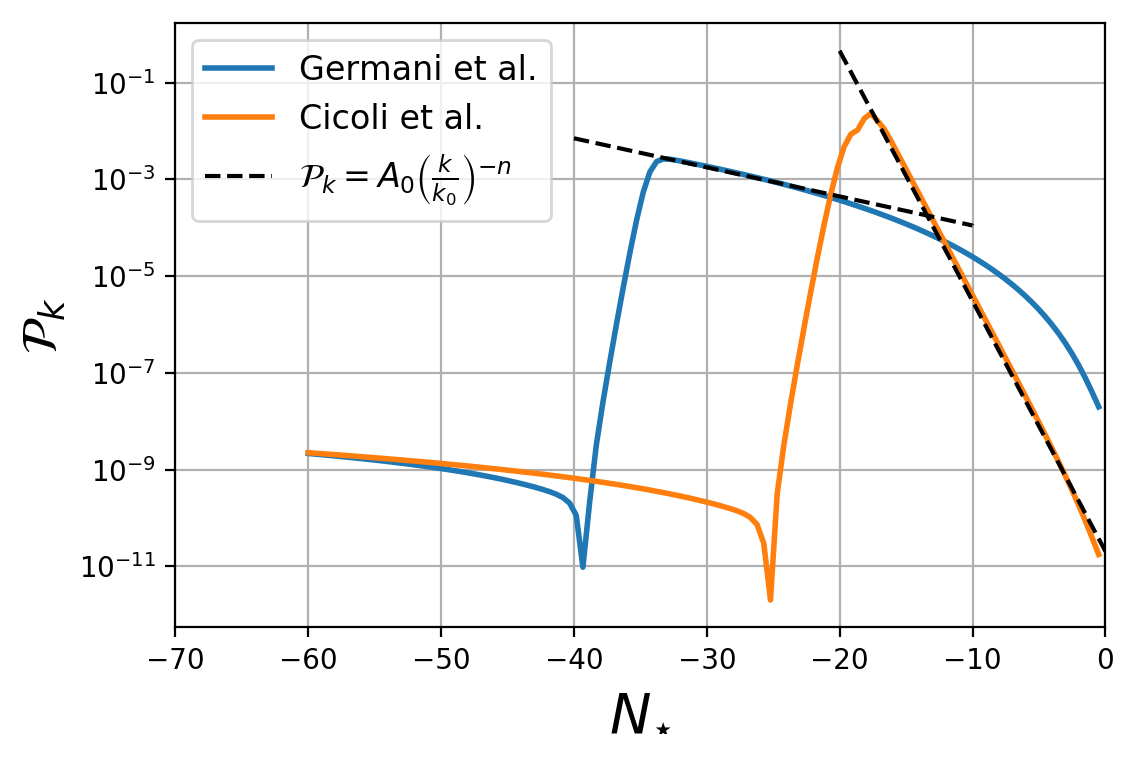}
\caption{Two typical examples of peaked power spectra. We see that the model (\ref{eq:model_pw}) provides a good description on how the power spectrum decays at large momenta.}
\label{fig:PP_vs_model}
\end{center}
\end{figure}

Close to the peak, the curvature power spectrum defined as, see e.g. \cite{ilia}, 
\be
{\cal P}(k)\simeq a^4 H^4\frac{81}{16\times 2\pi^2}\frac{P_{\Delta}(k,t)}{k}\ ,
\ee
is well described, in all models discussed so far, by the following template 
\begin{equation}
  \P(k) = 
   \begin{cases}
    0 & \text{for } k < k_{\text{peak}} \\
  \P_0 \left(\frac{k}{k_{\text{peak}}}\right)^{-n} \ , & \text{for } k \ge k_{\text{peak}}\ .
   \end{cases}\label{eq:model_pw}
\end{equation}
In \eqref{eq:model_pw}, the spectral index $n$ is standardly related, in the limit $\epsilon_V\ll \eta_V$, to the second derivative of the potential at the maximum via $n\simeq-3+\sqrt{9-12\eta_V}$. In particular, all power spectrum considered in the literature so far decay at a rate that goes from $n\sim0.15$ to $n\sim 2.7$. In Fig. \ref{fig:PP_vs_model} we show two examples and corroborate that (\ref{eq:model_pw}) provides an accurate description of how the two-point function decays after the peak.  

For simplicity, in this section we use a hard cutoff at small scales rather than a gaussian, and thus compute the variance and skewness integrating from $k=0$ to $k=k_{\rm cut}$.

\subsection{Predictions}

With the power spectrum template introduced before we find 
\be\label{nu_temp}
\nu_c=\frac{9\,\F_0^c\, r_{\rm m}^2}{4\, \sqrt{\P_0}\, (k_{\text{cut}}r_m)^2}\left(\frac{4-n}{\gamma^{n-4}-1}\right)^{1/2} \ ,
\ee
\noindent where $\gamma\equiv k_{\rm peak}/k_{\rm cut}$. By considering the non-gaussianities to be of local shape, as in all model studied so far, $E_3$ can be written as
\be\label{E3_gen}
 E_3 = f(n,\gamma,\P_0) \,f_{\rm NL} \, \nu^3 
 \ ,
\ee
\noindent where the function $f(n,\gamma,\P_0)$ is
\be
f(n,\gamma,\P_0) \simeq \sqrt{\P_0}\left(\frac{n-4}{n-2}\right)\left(\frac{\gamma^n-\gamma^4}{4-n}\right)^{1/2}\frac{\left(\gamma^6+\gamma^{2n}-\gamma^{2+n}-\gamma^{4+n}\right)}{\left(\gamma^4-\gamma^n\right)^2} \ .
\ee\label{eq:fE3_gen}
For a given $n$ and a fixed amplitude of the power spectrum, $\nu_c$ is a function of $\gamma$ with a minimum $\nu^{\rm min}$ at certain scale $k_{\rm cut}^{\rm min}$. Since the abundances are exponentially suppressed by $\nu_c$, the desired value of $\nu_c$ must be precisely $\nu^{\rm min}$.


For $n=0$ to $n=2$, in order to have the typical range $\nu_c=[5,10]$ we find  $10^{-3}\lesssim\,\P_0\lesssim 10^{-2}$
\be
f_{\rm NL} \lesssim  10^{-2} \ ,
\ee
while the constraint gets relaxed for $n=3$: $10^{-2}\lesssim\,\P_0\lesssim 10^{-1}$ and $f_{\rm NL} \lesssim  10^{-1}$.

Such bounds for $f_{\rm NL}$ are respected in slow-roll inflation when the slow-roll parameters are small. However, in transient USR/CR these parameters could be large and so the bounds on $f_{\rm NL}$ could be violated. As we have shown, this indeed happens in all current inflationary models related to PBHs formation.

\section{Conclusions}\label{sec:conclusions}

By the use of peak theory, in this paper we have re-analysed all current models of single-field inflation able to produce large peaks in the power spectrum of curvature perturbations. In particular, we have shown that models featuring a maximum in the potential could match the required abundances of PBHs according to a gaussian statistics of curvature perturbations. However, in those cases, we found that non-gaussianities are large enough to spoil the gaussian predictions. Whenever the peak is instead generated by a transient ultra-slow-roll trajectory, we confirm that the current available model cannot match the necessary abundances, given the relatively small amplitude of the power spectrum. This case is nevertheless interesting as the value of $f_{\rm NL}$ is here smaller than in the transient CR trajectory. Thus, an inflationary model with transient USR phase might be the one to look at in order to predict the right PBHs abundances by the sole use of gaussian statistics.

Interestingly, we have also shown that all known models producing a peak in the power spectrum fall into a single class where the peak happens very close, in field space, to the local maximum of the potential. In these cases, we proved that the non-gaussianity at the peak of the power spectrum is of the local shape and its amplitude is related to the slow-roll parameters at the beginning of the graceful exit of inflation ($\epsilon_2^{ge}$) by the relation 
\begin{align}
f_{\text{NL}}\Big |_{\rm at\ the\ peak}&\simeq \frac{5}{12}\epsilon_2^{ge}\ . \nonumber
\end{align}

\setcounter{equation}{0}
\section*{Acknowledgments} 
We thank Jaume Garriga, Ilia Musco and Toni Riotto for many discussions. CG wish to thank P.S. Corasaniti for discussions on windows functions. VA is supported by FPA2016-76005 -C2-2-P, MDM-2014-0369 of ICCUB (Unidad de Excelencia Maria de Maeztu), AGAUR2014-SGR-1474 and AGAUR2017-SGR-754. CG is supported by the Ramon y Cajal program and partially supported by the Unidad de Excelencia Mar\'ia de Maeztu Grant No. MDM-2014-0369 and by the national FPA2013-46570-C2-2-P and FPA2016-76005-C2-2-P grants.

\end{document}